\documentclass[pre,twocolumn,twoside,superscriptaddress,showpacs,floatfix]{revtex4-2}
\usepackage{amssymb,amsmath,graphicx,multirow,natbib,hyperref,color}
\hypersetup{colorlinks,citecolor=black,filecolor=black,linkcolor=black,urlcolor=black}

\begin{document}
\title[]{Coupling disorder in a population of swarmalators}
\author{Hyunsuk \surname{Hong}}
\affiliation{Department of Physics, Jeonbuk National University, Jeonju 54896, Korea}
\affiliation{Research Institute of Physics and Chemistry, Jeonbuk National University, Jeonju 54896, Korea}
\author{Kangmo \surname{Yeo}} 
\affiliation{Department of Physics, Jeonbuk National University, Jeonju 54896, Korea}
\author{Hyun Keun \surname{Lee}} 
\affiliation{Department of Physics, Sungkyunkwan University, Suwon 16419, Korea}
\date{\today}
\begin{abstract}
We consider a population of two-dimensional oscillators with random couplings, and explore the collective states. The coupling strength between oscillators is randomly quenched with two values one of which is positive while the other is negative, and the oscillators can spatially {\it{move}} depending on the state variables for phase and position. We find that the system shows the phase transition from the incoherent state to the fully synchronized one at a proper ratio of the number of positive couplings to the total. The threshold is numerically measured, and analytically predicted by the linear stability analysis of the fully synchronized state. It is found that the random couplings induces the long-term state patterns appearing for constant strength. The oscillators move to the places where the randomly quenched couplings work as if annealed. We further observe that the system with mixed randomnesses for quenched couplings shows the combination of the deformed patterns understandable with each annealed averages.
\end{abstract}



\maketitle

\section{\label{intro} Introduction}
In recent studies~\cite{Kevin17,Hong18}, oscillators that sync and swarm, called {\it{swarmalators}}, have been considered, and interesting long-term states have been found. The swarmalators can move in the space under the correlated dynamics of the phase and space variables of each oscillator. The {\it{mobile}} feature of the oscillators is found to induce nonstationary states such as the active-phase wave state and the splintered phase wave state. The collective property of finite number of swarmalators is studied~\cite{Kevin18}, the implementation in robots is tried as an application~\cite{Bett20}, and various steady states by finite interaction-distance is reported~\cite{Lee21}. Mathematical property~\cite{swm2}, model extension~\cite{swm4,swm5,swm6}, and minimal modeling~\cite{swm7} are of basic interest. Application to swarming and flocking of biological colony~\cite{swm} is a primary interest. Control of engineering objects like drones or robots is also an application front~\cite{swm3}.

In the study of swarmalators, the interesting incoherent states are usually the consequence of negative-definite coupling strength in phase dynamics (the positive-definite case is also in Ref.~\cite{Hong18} where noise is added). On the other hand, the coupling characteristic among the constituents of many systems is rather complex and not so simple. For example, the interaction among the neurons in the neural network systems is given by the mixture with positive and negative ones~\cite{neuron,neuron2}, in general. As another example, Japanese frogs' calling behavior can be understood by considering the mixture of positive and negative interaction with each other~\cite{Ikkyu08}. Probably, the spin glass ~\cite{sg,spinglassXY}, to which the various interesting properties of condensed matter is attributed, can be the representative example for the non-definite sign of the couplings. Considering those features in nature, the mixed coupling with positive and negative strength deserves to be considered for understanding the collective behavior of the real systems, which motivates the present study.

In this paper, we consider a population of the oscillators that can sync and swarm, governed by the random interaction in the phase dynamics. We explore how the coupling-disorder affects the long-term states in the system. In particular, we pay attention to the possibility of the phase transition~\cite{ptStanely} in the system, and focus on whether the patterns observed in the absence of coupling disorder still appear. The effective annealing of the randomly quenched coupling strengths to their average is suggested in the mobility of swarmalators to understand the numerical results. Phase transition to the fully synchronized phase, reproducibility of the phases known in the original model of the no coupling disorder, and mixture of deformed long-term states by coupling disorders are explained from the viewpoint of the annealed couplings.

This paper is organized as follows: Section~\ref{mdl} introduces the model of coupled swarmalators with random couplings, and Sec.~\ref{pt} shows the collective behavior and the phase transition in the system. In Sec.~\ref{ls} we derive the threshold of the transition by the linear stability analysis of the fully synchronized state. Various long-term states including the nonstationary ones are shown in  Sec.~\ref{nonss}, and the system with couplings of more than one quenched randomness are understood as the combination of annealed-coupling systems in Sec.~\ref{mrc}. A brief summary is given in Sec.~\ref{sum}.

\section{\label{mdl} Model}
The generalized model of $N$-coupled oscillators that we consider here is given by
\begin{eqnarray}
\frac{d\theta_i}{dt} &=& 
\frac{1}{N}\sum_{j\neq i} F(r_{ij}) K_{ij} \sin(\theta_j-\theta_i),
\label{eq:model_theta} \\
\frac{d{\bf{r}}_i}{dt} &=& 
\frac{1}{N}\sum_{j\neq i}
G(r_{ij}, \theta_{ij})
\label{eq:model_r}
\end{eqnarray}
for $i=1,\cdots,N$, where $\theta_i$ and ${\bf{r}}_i$ represent the phase $(0 \leq \theta_i \leq 2\pi)$ and the position vector of the $i$th oscillator, respectively. $F(r_{ij})$ is a function for the spatial dynamics of the oscillators with $r_{ij}=|{\bf{r}}_{ij}|$ for ${\bf{r}}_{ij}\equiv{\bf{r}}_j - {\bf{r}}_i$, and $G(r_{ij},\theta_{ij})$ is a function for spatial dynamics with $\theta_{ij}\equiv\theta_j - \theta_i$. $K_{ij}$ denotes the random coupling strength between the oscillators $i$ and $j$, having the symmetric property $K_{ij}=K_{ji}$. It represents a sort of {\it{bond}} coupling between the oscillators, instead of the {\it{site}} one ($K_i$) that affects the oscillator itself. For simplicity, we randomly chose $K_{ij}$ from the two-peaks distribution
\begin{equation}
h(K_{ij})=p\delta(K_{ij}-K_p)+(1-p)\delta(K_{ij}-K_n),
\label{eq:hK}
\end{equation}
where $p$ is the probability of the positive coupling, and $K_{p} > 0$ and $K_{n} < 0$. For convenience, the ratio $Q\equiv -K_{n}/K_{p} (>0)$ has been chosen for the control parameter. With the functions $F$ and $G$ in the model, the phase variable $\theta$ and the spatial one $r$ become correlated, which means the oscillators can move around in the space. Such correlation in the dynamics of the space and the phase of the oscillators has been also studied in Ref.~\cite{Kevin17}, where the authors found five long-term states including the two nonstationary ones. Note that the phase coupling in Ref.~\cite{Kevin17} does not have any disorder. We here consider coupling disorder that is given by the random values from the distribution in Eq.~\eqref{eq:hK}.

We notice that the special case with $F(r_{ij})=1$ and $G(r_{ij},\theta_{ij})=0$ is equivalent to the mean-field $XY$ model~\cite{spinglassXY} with {\it{random}} coupling strength, governed by the Hamiltonian ${\cal{H}}=-\frac{1}{2N}\sum_{i\neq j} K_{ij}\cos(\theta_j - \theta_i)$. Equation~(\ref{eq:model_theta}) with $F=1$ and $G=0$ leads to the overdamped version of the Hamiltonian dynamics at zero temperature, and interestingly it is found that the first-order phase transition occurs at the same threshold $p_c$ as that we study here~\cite{Hong21}.

In this paper, we consider $F=1/r_{ij}$. The function $G$ consists of the attraction and repulsion forces acting on each oscillator, where the force functions are taken as the algebraic ones with a power like $r_{ij}^{-a}$ following Ref.~\cite{Kevin17}. The model is then given by 
\begin{eqnarray}
{\dot{\theta}}_{i} &=& \frac{1}{N}
\sum_{j\neq i} 
\frac{K_{ij}}{r_{ij}}\sin(\theta_j-\theta_i), \label{eq:theta} \\
\dot{\bf{r}}_i &=& \frac{1}{N}\sum_{j\neq i} 
\left[\frac{{\bf{r}}_{ij}}{r_{ij}}(A+J\cos(\theta_j-\theta_i)) - B \frac{{\bf{r}}_{ij}}{r^2_{ij}}\right], \label{eq:x}  
\end{eqnarray}
where $A$ and $B$ are respectively the parameters for the attractive force and the repulsive force. Here, we choose $A=B=1$ for convenience. $J$ is the parameter which measures how the phase similarity enhances the spatial proximity.  For example, a positive value of $J$ means {\textit{like attracts like}}, i.e., the swarmalators tend to be near the other swarmalators with similar phases. On the other hand, a negative value of $J$ means the opposite case: The swarmalators prefer to be near the others with opposite phases. To keep the attractive force always positive, the values of $J$ are constrained like $-1 \leq J \leq 1$, following the Ref.~\cite{Kevin17}. The difference from the model of Ref.~\cite{Kevin17} is the generalization of the constant $K$ to pair-dependent $K_{ij}$. For $K_{ij} > 0$, the swarmalators prefer to have the similar phase, but for $K_{ij} < 0$, the opposite tendency occurs. We are interested in the effect of random $K_{ij}$ in the formation of long-term states.

\section{\label{pt} Phase transition}
\begin{figure}
\includegraphics[width=9.0cm]{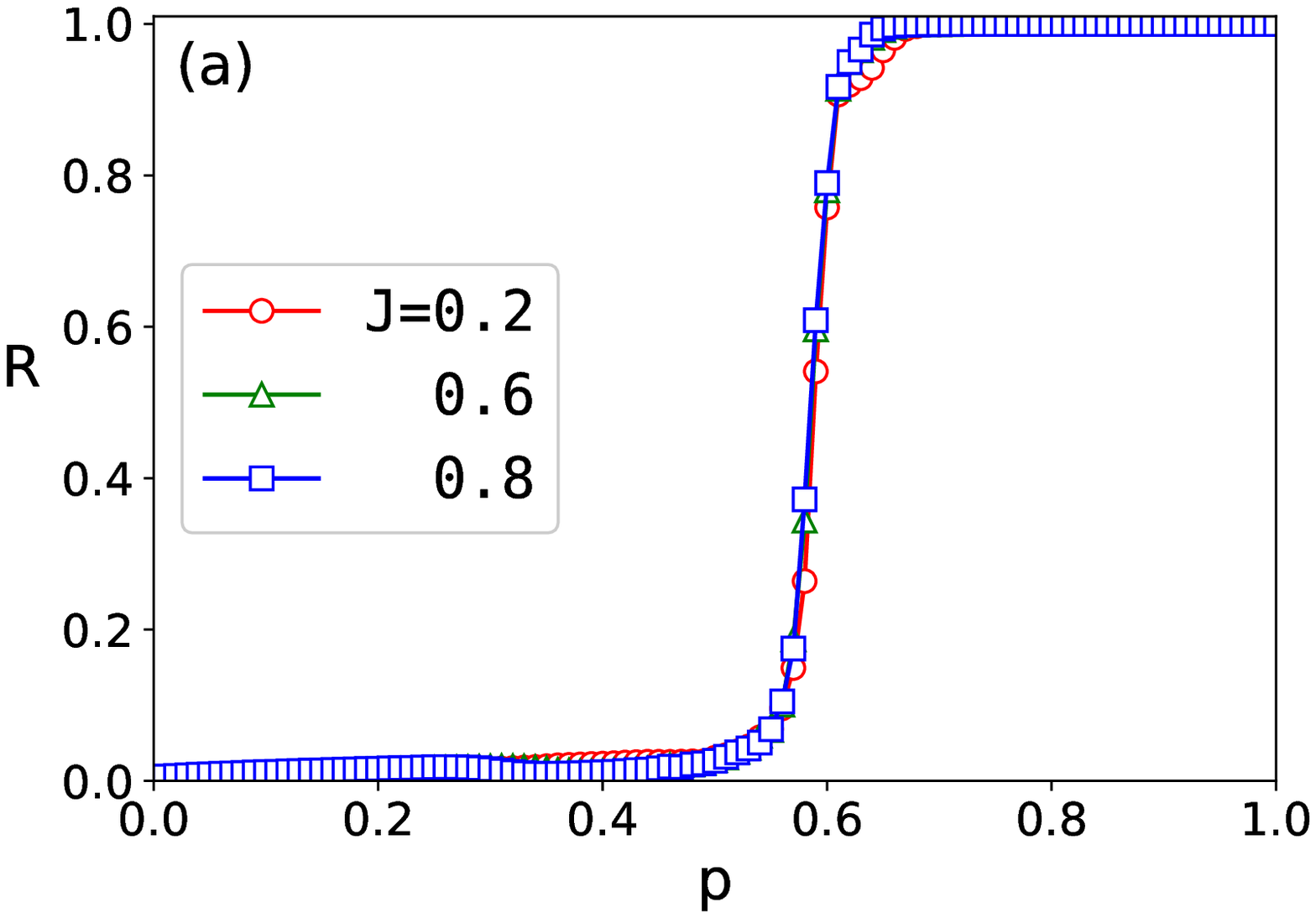}
\includegraphics[width=9.0cm]{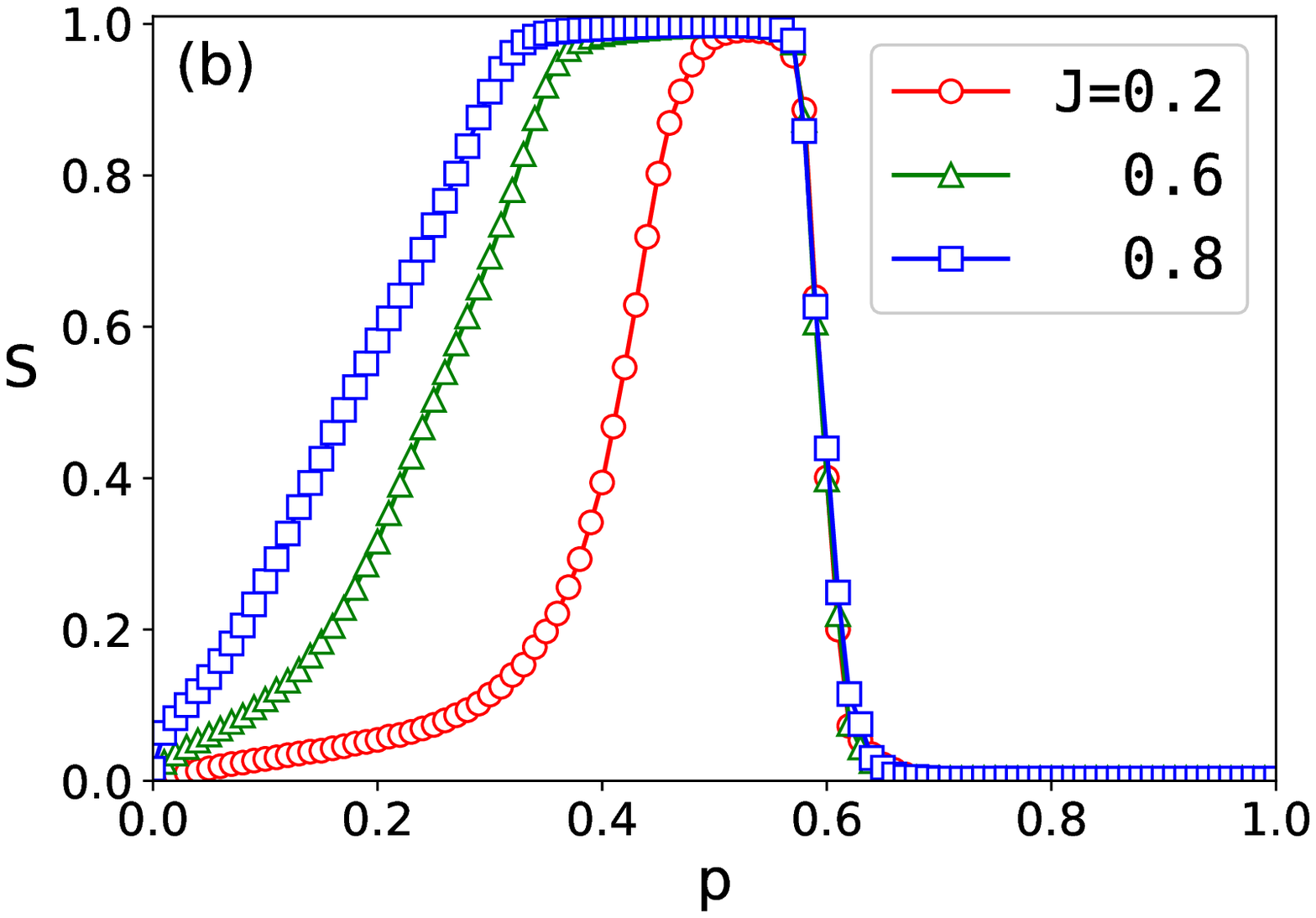}
\caption{(Color Online)
(a) $R$ is shown as a function of $p$ for various values of $J$; (b) $S$ vs. $p$ for various values of $J$. Parameters: $Q=1.25$ and $N=200$. 
}
\label{fig:RS}
\end{figure}
To see the collective states in the system and the effects of random coupling on the states we perform the numerical simulations on Eqs.~(\ref{eq:theta}) and (\ref{eq:x}). We first investigate the phase synchronization behavior of the oscillators, by measuring the complex order parameter defined by~\cite{Kuramoto84} 
\begin{equation}
Re^{i\Theta} = \frac{1}{N} \sum_{j=1}^N e^{i\theta_j}, 
\label{eq:R}
\end{equation}
where $R$ measures the phase coherence, and $\Theta$ is the mean phase angle of the oscillators. For example, $R=0$ means the incoherent state where the all oscillators have the random phase $\theta_i\in[0,2\pi)$, and $R=1$ is the fully synchronized state where they all have the same one $\theta_j=\theta_s$ for all $j$'s.

Also, to see the correlation between the phase angle $\theta_j$ and the azimutal one $\phi_j=\tan^{-1}(y_j/x_j)$ for the $j$th oscillator, which is induced by the mobile feature of the oscillators in the system, we measure another order parameter defined by~\cite{Kevin17}
\begin{equation}
S_{\pm}e^{i\Psi_{\pm}} = \frac{1}{N} \sum_{j=1}^N e^{i(\phi_j \pm \theta_j)}, 
\label{eq:S}
\end{equation}
where $S_{\pm}$ is the magnitude, and $\Psi_{\pm}$ is its mean phase. Here, the order parameter $S$ measures the ``correlation'' between the spatial information (by the $\phi_j$) and the phase information (by the $\theta_j$). The system exhibits ``assortative and/or commutative'' correlation between the two depending on the initial conditions, where the $S_{+}$ comes from the assortative correlation, and $S_{-}$ comes from the commutative correlation. We chose $S$ by taking the maximum value of $S_{+}$ and $S_{-}$, i.e., $S={\rm{max}(S_{\pm})}$~\cite{Kevin17}.

In order to examine $R$ and $S$, we basically performed the numerical simulations using the Python or C programs. The total $5\times 10^4$ time steps with the discrete time unit $dt=0.1$ have been considered, where the first $2.5\times 10^4$ steps were discarded for the equilibrium, and then the time average of the quantities has been done for the later steps. And, the ten samples have been used for the average. Initial phases and positions were, respectively, randomly sampled from the uniform distributions over $[0,2\pi)$ and $[-1,1)\times[-1,1)$.

Figure~\ref{fig:RS} shows the behavior of $R$ and $S$ as a function of $p$ for various values of $J$. The value of $Q=1.25$ was chosen, and $K_p=1$ and $K_n=-1.25$ were chosen for convenience. We find that the system shows the phase transition behavior from the incoherent state ($R=0$) to the fully synchronized one ($R=1$) at a finite value of $p$, i.e., $p_c~\sim 0.6$. Interestingly, we find that the behavior of $R$ and its threshold $p_c$ do not depend on $J$, which can be easily understood by the absence of $J$ in the phase dynamics given by Eq.~(\ref{eq:theta}). On the other hand, the behavior of $S$ is different from that of $R$ as shown in Fig.~\ref{fig:RS}: In the regime of the fully synchronized state ($R=1$) for $p> p_c$, the order parameter $S$ is zero, which implies no correlation between the phase angle and the spatial angle. However, in the incoherent state with $R=0$ for $p<p_c$, $S$ shows finite value, which means there is some correlation between the phase angle and the spatial angle. Moreover, the value of $S$ for $p<p_c$ shows $S=1$ for a certain region of $p$, and the range of $p$ of this region depends on the value of $J$ as shown in Fig.~\ref{fig:RS}~(b): The range gets wider as $J$ increases.

\begin{figure}
\includegraphics[width=9cm]{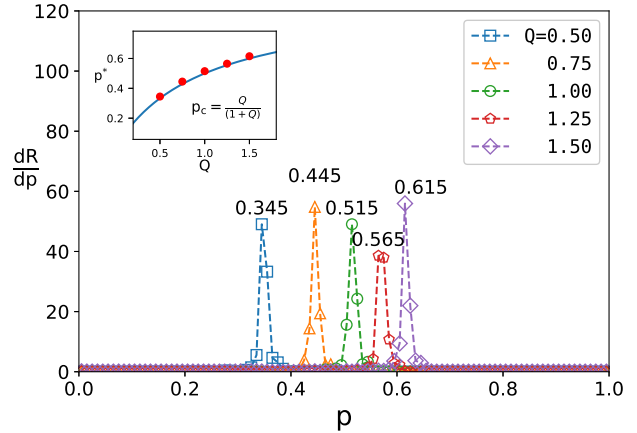}
\caption{(Color Online)
The first derivative of $R$ over $p$, $dR/dp$, is shown as a function of $p$ for various values of $Q$, where the value of $p$ at which $dR/dp$ reaches the maximum value is denoted as $p^*$. Inset: $p^*$ vs. $Q$ is shown, where the blue solid line is the theoretical prediction given by $p_c=Q/(1+Q)$, and the value of $p^*$ is consistent with the prediction. The system size is $N=800$ and $J=0.8$, and the data have been averaged over 10 samples.
}
\label{fig:pc_Q}
\end{figure}

To pin down the threshold $p_c$, we now estimate it numerically. We measure the value of $p$ at which the first derivative $dR/dp$ shows the maximum. Figure~\ref{fig:pc_Q} shows the behavior of $dR/dp$ as a function of $p$ for various values of $Q$, whereas the value of $p$ at which the $dR/dp$ reaches the maximum is denoted as $p^*$. The inset of Fig.~\ref{fig:pc_Q} shows $p^*$ as a function of $Q$, which shows a good consistency with the theoretical prediction $p_c=Q/(1+Q)$ (see Sec.~\ref{ls}). 

\section{\label{ls} Linear stability of the fully synchronized state ($R=1$)}
In this section, we investigate the linear stability of the fully synchronized state with $R=1$. Let a fully synchronized state have a common phase $\theta_i=\theta_{\rm s}$ for all $i$. In order to examine its linear stability, we consider a slightly perturbed situation such as $\theta_i =\theta_{\rm s} + \phi_i$ with small perturbation $\phi_i$ for all $i$. For this setting, Eq.~\eqref{eq:theta} is linearized to
\begin{equation}
\label{lt}
\dot \phi_i 
= \frac{1}{N}\sum_{j\neq i} \frac{K_{ij}}{r_{ij}}(\phi_j-\phi_i)\,.
\end{equation}
Separating the summation according to the values of $K_{ij}$, we rewrite Eq.~\eqref{lt} as
\begin{eqnarray}
\label{lt2}
\dot \phi_i = \frac{1}{N}
&&\left( K_p \sum_{j\in C_p(i)} \frac{1}{r_{ij}}(\phi_j-\phi_i) \right. 
\nonumber \\
&& \left.~~~+K_n \sum_{j\in C_n(i)} \frac{1}{r_{ij}}(\phi_j-\phi_i)\right),
\end{eqnarray}
where $C_p(i)$ is the collection of $j$ having $K_{ij}=K_p$ and $C_n(i)$ is that for $K_{ij}=K_n$.

We here note that $K_{ij}$ is not included in the spatial dynamics [see Eq.~\eqref{eq:x}]. Thus, for a fixed $i$, the fraction of such $j$'s giving $K_{ij}=K_p$ in the total $N$ would be approximately $p$, where the deviation is expected to decrease as $N$ increases. Similarly, that for $K_{ij}=K_n$ is given by $(1-p)$. This consideration motivates us to replace $\sum_{j\in C_p(i)}$ and $\sum_{j\in C_n(i)}$ with $p\sum_{j(\neq i)}$ and $(1-p)\sum_{j(\neq i)}$, respectively, in Eq.~\eqref{lt2} when $N$ is large enough. Therefore, for large $N$, Eq.~\eqref{lt2} can be cast into
\begin{equation}
\label{lt3}
\dot \phi_i = 
\langle K_{ij}\rangle 
\frac{1}{N}\sum_{j\neq i} \frac{1}{r_{ij}}(\phi_j-\phi_i)\,.
\end{equation}
where $\langle K_{ij}\rangle\equiv pK_p +(1-p)K_n $ is the average of $K_{ij}$ for the two-peak distribution $h(K_{ij})$.

Introducing $m_{ij} = -1/(Nr_{ij})$ for $i\neq j$ and $m_{ii}=-\sum_{j\neq i}m_{ij}$, one writes Eq.~\eqref{lt3} as
\begin{equation}
\label{lt4}
\dot \phi_i = 
-\langle K_{ij}\rangle \sum_{j=1}^N m_{ij}\phi_j\,.
\end{equation}
Interestingly, the matrix $m=\{m_{ij}\}$ is positive-definite for large $N$ as follows. It holds $|m_{ij}| \sim {\cal O}(1/\sqrt{N})$ for off-diagonal terms. This is based on the numerical observation that all swarmalators reside in the two-dimensional spatial region of the ${\cal O}(1)$-area and there is no concentration of them. Thus, $r_{ij}\sim 1/\sqrt{N}$ holds, from which $|m_{ij}| \sim {\cal O}(1/\sqrt{N})$ follows. It also follows that $m_{ii}=-\sum_{j\neq i}m_{ij} \sim {\cal O}(\sqrt{N})$. Then, as $N$ increases, $m=\{m_{ij}\}$ approaches $\bar m = \{\bar m_{ij}\}$, which is given by
\begin{equation}
	\label{m-infty}
	\bar m_{ij}=\delta_{ij}\bar m_i\,,
\end{equation}
where $\delta_{ij}$ is the Kronecker delta that assigns $1$ when it assigns $i=j$ while $0$ otherwise. The matrix $\bar m$ is positive-definite because $\bar m_i\equiv\lim_{N\to\infty}m_{ii}>0$. Therefore, $m$ is also positive-definite for sufficiently large $N$.

When the positive definiteness of $m$ is considered in Eq.~\eqref{lt4}, one knows that the sign of $\langle K_{ij}\rangle$ determines whether $\phi_i$ will decay or not. This is because every eigenvalue of the positive-definite matrix is positive. Thus, $\phi_i$ decays when $\langle K_{ij}\rangle>0$, otherwise it grows when $\langle K_{ij}\rangle<0$. Since $Q\equiv -K_n/K_p$, we get $\langle K_{ij}\rangle =  K_p(p+(p-1)Q)$. Therefore, $\langle K_{ij}\rangle>0$ corresponds to $p>Q/(1+Q)$ and $\langle K_{ij}\rangle<0$ corresponds to $p<Q/(1+Q)$, which leads to  
\begin{equation}
\label{pc}
p_c=Q/(1+Q)\,.
\end{equation} 
For $Q=1.25$, used for the numerical data shown in Figs.~\ref{fig:RS} and \ref{fig:pc_Q}, Eq.~\eqref{pc} gives $p_c=5/9=0.55..\,$. The deviation from the numerical value $p_c\approx 0.6$ in the figures is supposed to be a finite $N$ effect. Note $p_c\approx 0.565$ of Fig.~\ref{fig:pc_Q} for $N=800$ is closer to the prediction $p_c = 5/9$ than the prediction in Fig.~\ref{fig:RS} for $N=200$. It is interesting that the threshold value shown in Eq.~(\ref{pc}) is equivalent to that for the mean-field $XY$-type oscillators with random coupling disorder~\cite{Hong21}. This means that the spatial dynamics in the system does not change the threshold value of the phase transition from the incoherent state to the fully synchronized one.

The crucial step of our argument above is the use of $\sum_{j\in C_p(i)}=p\sum_{j(\neq i)}$ and $\sum_{j\in C_n(i)}=(1-p)\sum_{j(\neq i)}$ to write Eq.~\eqref{lt3}. We here remark this is basically an annealed approximation. In fact, $K_{ij}$ is quenched by definition in the model and has never been treated in an annealed way in the numerical work. Thus the consistency between Eq.~\eqref{pc} and its numerical counterpart is interesting. Equation~\eqref{eq:x} gives a clue to understanding how the quenched $K_{ij}$ works as if annealed. The spatial dynamics governed by Eq.~\eqref{eq:x} is indifferent to $K_{ij}$. Based on this, one may expect the distribution of $K_{ij}$ for given $i$ has no position-dependence, i.e., the fractions for $K_p$ and $K_n$ are, respectively, $p$ and $(1-p)$ independent of position. This results in the annealed average of phase dynamics in Eq.~\eqref{eq:theta}. The consistency between the numerical data and the annealed average, even in the presence of quenched $K_{ij}$, means that the oscillators move to the proper places. That is, the effective annealing of $K_{ij}$ to $\langle K_{ij}\rangle$ is a characteristic property of swarmalator that can move.

We also examine that the sign of $\langle K_{ij}\rangle$ always governs the onset of the sync state in further numerical study using other $h(K_{ij})$ distributions. These observations suggest that the annealed approximation could be generally available. We remark that the effective annealing requiring proper locations of swarmalators is considered for long-term states.

\section{\label{nonss} Non-stationary states in the incoherent regime ($R=0$)}
\begin{figure}
\includegraphics[width=9cm, height=8.5cm]{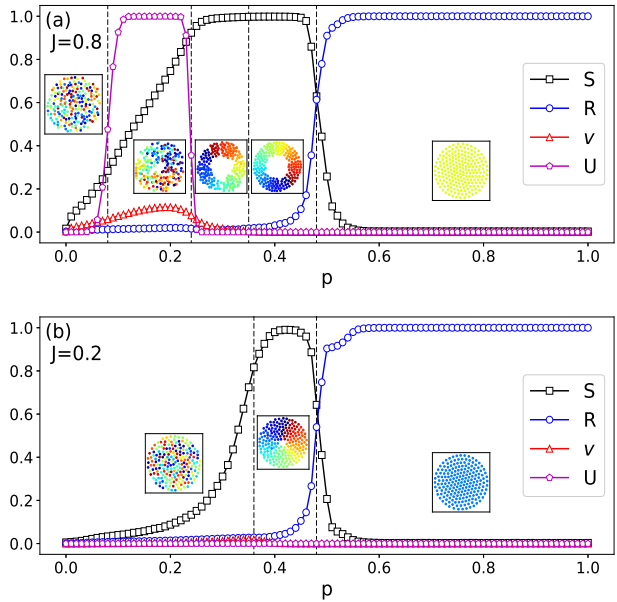}
\caption{(Color Online) Behavior of $S$, $R$, $V$, and $U$ is shown as a function of $p$ for (a) $J=0.8$ and (b) $J=0.2$. Snapshots for the corresponding regime are also shown in the insets.  Parameters: $N=200$ and $Q=0.8$.}
\label{fig:RSUV}
\end{figure}
In this section, we investigate the incoherent state $(R=0)$ further. In particular, we pay attention to the possibility of the nonstationary states such as the active phase wave (APW) state and the splintered phase wave (SPW) state found in Ref.~\cite{Kevin17}. Therein, the all coupling strength among the oscillators was chosen as the negative one with no disorder like $K_{ij}=K$, which corresponds to $p=0$ in the current study. We note that, in Ref.~\cite{Kevin17}, various long-term states such as the async state, the APW state, the SPW state, and the static phase wave state were found depending on the strength of the negative coupling in the phase dynamics. Especially, the APW state and the SPW state are the nonstationary states where the oscillators are not static in both phase and space. It is thus interesting whether the dynamics of Eqs.~\eqref{eq:theta} and \eqref{eq:x} with quenched disorder $K_{ij}$ can induce such a nonstationary feature. Below, we again resort to the numerical data by the same scheme used for Fig.~\ref{fig:RS}.

To see the possibility of the nonstationary states for $p>0$, we measure the following two quantities. One is the mean velocity $V$ defined by~\cite{Kevin17,Hong18}
\begin{equation}
V = \frac{1}{N}\sum_{j=1}^N v_i, 
\label{eq:V}
\end{equation}
where $v_i = \sqrt{{{\dot{x}}_i}^2+{{\dot{y}}_i}^2}$. The mean velocity measures the nonstatic property of the state. In other words, the finite value of $V$ means that the swarmalators move around in the plane in the long-term state. The other one is~\cite{Kevin17,Hong18} 
\begin{equation}
U = \frac{N_{\rm{rot}}}{N}, 
\label{eq:U}
\end{equation}
where $N_{\rm{rot}}$ is the number of swarmalators circulating at least $2\pi$, and thus $U$ represents the fraction over the total swarmalators~\cite{Kevin17,Hong18}. Note that $U$ can play as the role of the indicator for the active phase wave state~\cite{Kevin17}. Figure~\ref{fig:RSUV}(a) and (b) show the quantities $R$, $S$, $V$, and $U$ as a function of $p$ for $J=0.8$ and $0.2$, respectively. As $p$ increases from zero, we find that the phase transition occurs from the incoherent state $(R=0)$ to the fully synchronized state ($R=1$) at a finite value of $p_c$, as shown in the behavior of $R$. Interestingly, the value of $p_c$ for two different values of $J$ are found to be same, which means the threshold does not depend on $J$. On the other hand, the behavior of $S$, $U$, and $V$ is found to differ depending on the value of $J$ (See Fig.~\ref{fig:RSUV}).

When $R=0$, note that the system shows a finite value of $S$ through the splintered and static phase wave. Also, the system shows the active phase wave state for small $p$ when $J$ is large.  This active phase wave does not occur when $J$ is small. We find that the system shows both the stationary states and the nonstationary states depending on $p$ for a given value of $Q$ and $J$.  Especially, when $J$ is large $(J=0.8)$, all five long-term states that have been reported in Ref.~\cite{Kevin17} exist depending on $p$.  It is interesting to note that the nonstationary states such as active phase wave sate and splintered phase wave also exist in the presence of the positive coupling among the swarmalators in the system ($p>0$).

\section{\label{mrc} Mixture of random couplings}
Above, we have seen that the result by random $K_{ij}$ is comparable with that by constant $K=K_{ij}$. The long-term state patterns, including the fully synchronized one, from both settings are similar for large $J$. It is remarkable that such analogy holds in the presence of finite faction for negative $K_{ij}$ that may bring about the frustration~\cite{frst,frst2} in the phase dynamics. This strongly suggests the annealed property conjectured for the onset of the sync phase at $\langle K_{ij} \rangle =0^+$ is also valid for non-zero $\langle K_{ij} \rangle$ cases. This motivates us to test the long-term states in the system with such quenched random couplings by mixture of different randomnesses. We are interested in whether the characters of long-term states attributed to each randomnesses will last or not.

Consider $N_1$ number of swarmalators in total $N$, and call them group 1 (G1). The coupling strength $K_{ij}$, therein, is assigned with $1$ for probability $p_1$ or $-1$ for $1-p_1$. Next, consider the other group, G2, of $N_2(=N-N_1)$ swarmalators, where $K_{ij}=\pm 1$ is similarly assigned using probability $p_2$. We also consider coupling between the swarmalators in both groups as assigning $1$ to $K_{ij}$ for probability $p_{\rm m}$ or $-1$ for $1-p_{\rm m}$. Below, we use $p_{\rm m}=p_2$ for simplicity. When $(N_1,p_1)=(100,1)$ and $(N_2,p_2)=(100,0)$, one may expect the sync disc in G1 and the APW state in G2 if the couplings that work between the groups do not overwhelm each character. Note the average coupling in G2, $\langle K_{ij} \rangle_2=-1$, is in the region for the APW state in the phase diagram reported in Ref.~\cite{Kevin17}. We obtained numerical data with time unit $dt=0.01$ and consider $2\times 10^5$ time steps.

\begin{figure}
\includegraphics[width=9.0cm]{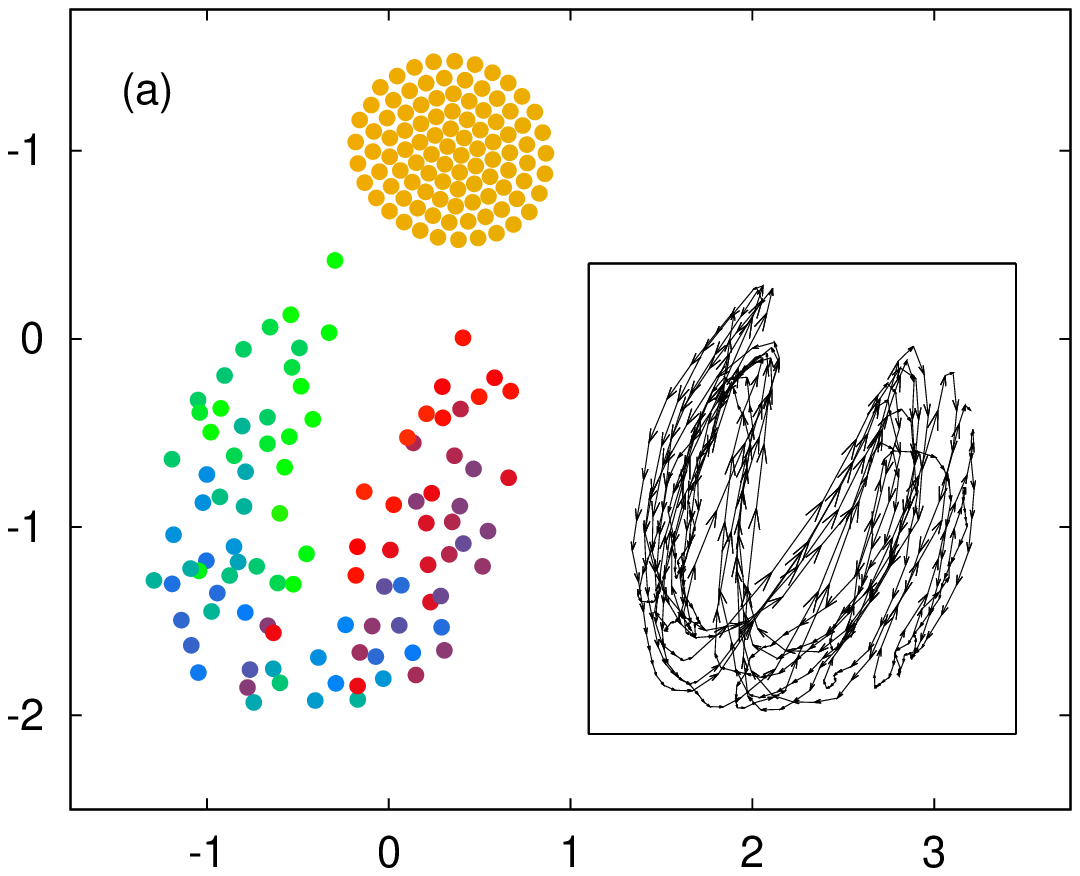}
\includegraphics[width=9.0cm]{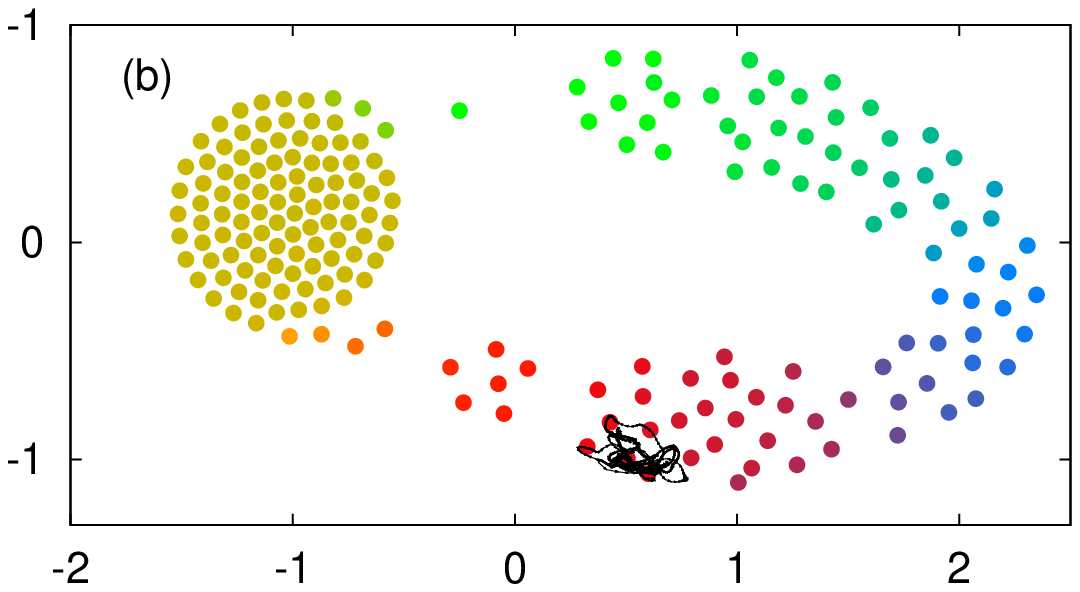}
\caption{(Color Online)
(a) Long-term state pattern for $N_1=N_2=100$, $p_1=1$, $p_2=0$, $p_{\rm m}=p_2$, and $J=1$. The inset is the trajectory of an arbitrarily chosen swarmalator in the U-shaped region, which sweeps the region. (b) Long-term state pattern for the same setting used in panel (a) except $p_2=0.4$. The black solid curve is the trajectory of a swarmalator localized there.
}
\label{fig-mix}
\end{figure}

Figure~\ref{fig-mix}(a) shows the numerical result. 
Two patterns appear; one is a slightly deformed sync disc while the other is U-shaped one. We have observed the former vibrates a little (not demonstrated here) and the latter is an active phase state. The inset in Fig.\ref{fig-mix}(a) is the trajectory of a swarm in the U-shape. Since it sweeps the region while changing the phase, we name the state the deformed APW (dAPW). It is interesting to note that this reminds us of the ``chimera state'' observed in the oscillators with identical frequency~\cite{chimera,chimera2,chimera3,chimera4}: The partial group of the swarmalators in the system exhibits the sync cluster, but the other group of the swarmalators do not join the sync cluster, showing the U-shape APW.

\begin{figure}
\includegraphics[width=9cm]{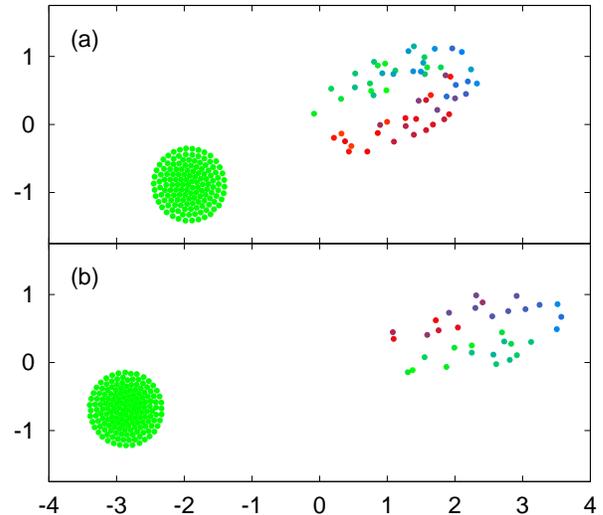}
\caption{(Color Online)
Long-term state patterns composed of sync disc and dAPW for $N_1=60$, $N_2=140$, $p_1=1$, $p_2=0$ in (a) and for $N_1=140$, $N_2=60$, $p_1=0.7$, $p_2=0$ in (b).
}
\label{fig-mix2}
\end{figure}

We vary $N_1$ to see a possible change in pattern. One easily see the distance between the sync disc and the dAPW increases when comparing Figs.~\ref{fig-mix}(a) and \ref{fig-mix2}(a). We also see the sync disc becomes more circular and static when the distance increases.
We next test whether a sync disc by $p_1<1$ in G1 only system ($N_2=0$) will still last after G2 is involved ($N_2>0$). The interest is a validity of the annealed approximation of $K_{ij}$ in G1. Figure~\ref{fig-mix2}(b) is the result for $p_1=0.7$, which is also composed of the sync disc and the dAPW. The sync disc is almost circular and static with enough separation from the dAPW.

We finally test whether G2 can show the SPW state while G1 remains a sync disc. As the splintered state appears for the negative coupling larger than that for APW, we use $p_2=0.4$, which corresponds to $\langle K_{ij}\rangle_2 = -0.2$, the average coupling strength in G2. This value is in the region for the splintered wave in the phase diagram reported in Ref.~\cite{Kevin17}. Figure~\ref{fig-mix}(b) is the result when the other settings are same as those for Fig.~\ref{fig-mix}(a). There still appear sync disc and U-shaped region. This time, the swarmalators of different colors in the U-shaped region are not mixed, different from the dAPW case. The gradual change of the swarmalators' colors in space is a characteristic of the splintered phase wave. The black solid curve near the bottom is the trajectory of a swarm, which remains near there as time goes on. As shown, it does not sweep the U region but localized in the small area while splintering. This is always the case for all swarmalators in the U-shaped region. We thus regard the U-shaped pattern as a deformed splintered phase wave (dSPW).

In the above numerical tests, the characters by each random couplings are preserved more or less in the mixed system. There apparently appear sync disc, APW, and splintered state though deformed. We here add the result is qualitatively similar for the various values of $p_{\rm m}$ (not shown here), and this does not depend on the initial condition. These observations suggest that the long-term states, demonstrated in Secs~\ref{ls} and \ref{nonss}, are still robust more or less in the mixed system. Interestingly, the patterns survive the frustration taking place between the subsystems G1 and G2. A mixture with async state or of more than two groups was not tested this time. Further study on various interesting properties of the mixed deformed patterns will appear elsewhere.

\section{\label{sum} Summary}
We considered the population of swarmalators with random coupling strength, and explored how the coupling disorder affects the long-term states in the system. In particular, the possibility of the phase transition and the robustness of the state patterns are focused. To understand the long-term states observed in the system with quenched disorder of coupling strength, we considered the effective annealed approximation, which is mediated by the mobility of swarmalators. In the viewpoint of annealed couplings, the numerical observation in the quenched system is explained and, furthermore, such a system of the mixture of different quenched disorders is also understood.

We found that the system shows the phase transition from the incoherent state to the fully synchronized state at a certain threshold $p_c$, where the value of $p_c$ is argued in the linear stability analysis of the fully synchronized state. We also found that, in the regime of the incoherent state below the threshold, various long-term states are found to exist. Especially, the nonstationary states such as the dAPW and the dSPW besides the normal APW and SPW are discovered.

All long-term states known for the bare model in Ref.~\cite{Kevin17} are realized in the system of random coupling strength. The values of average-coupling where each state appears are similar to their counterparts in the bare model. This strongly suggests the randomly quenched coupling strengths work as if annealed. The pattern by the supposed annealed coupling is so robust that the mixture of different random couplings leads to the proper combination of each deformed patterns.

\section{Acknowledgements}
This research was supported by NRF Grant No. 2021R1A2B5B01001951 and `Research Base Construction Fund Support Program' funded by Jeonbuk National University in 2021 (H.H), and by NRF Grant No. 2018R1D1A1B07049254 (H.K.L.).

\def\tb{\textbackslash}
 
\end{document}